\documentclass[a4paper]{article}
\usepackage {amssymb,latexsym}
\usepackage {amsmath}
\newtheorem{thm}{Theorem}
\newtheorem{prop}{Proposition}
\newtheorem{cor}{Corollary}
\newtheorem{lem}{Lemma}
\newtheorem{de}{Definition}

\newtheorem{rem}{Remark}
\newenvironment{proof}{
                        \noindent{\bf\small Proof: }\small}
                                       {\hfill {$\mathbf \Box$}\medskip}

\newcommand{\K}{\mathbb{K}}
\newcommand{\N}{\mathbb{N}}
\newcommand{\Z}{\mathbb{Z}}

\newcommand{\Hom}{\mathrm{Hom}}

\makeatletter
\@addtoreset{equation}{section}

\makeatother

\title{Note on the Cohomology of Color Hopf and Lie Algebras
}
\author{Xiao-Wu Chen\footnote{{\tt Supported by National Natural Science Foundation of
China (No.10501041) and AsiaLink project ``Algebras and
Representations in China and
Europe'' ASI/B7-301/98/679-11, xwchen@mail.ustc.edu.cn}.} \\
Department of Mathematics\\
University of Science and Technology of China\\
Hefei, Ahui, 230026, P.R.China\\
and\\
 USTC,
Shanghai Institute for Advanced studies\\
Shanghai, 201315, P. R. China\\
\\Toukaiddine Petit\footnote{{\tt Supported by the ESF
 Scientific Programme ``NOG'', toukaiddine.petit@ua.ac.be}. }
\\ Freddy Van Oystaeyen\footnote{{\tt Supported by the EC project
 Liegrits MCRTN 505078, fred.vanoystaeyen@ua.ac.be}.}
\\Department Wiskunde en Informatica
 Universiteit Antwerpen\\
B-2020, Antwerp, Belgium\\
}

\date{}

\begin{document}

\maketitle

\begin{abstract}
Let $A$ be a $(G, \chi)$-Hopf algebra with bijective antipode and
let $M$ be a $G$-graded $A$-bimodule. We prove  that there exists
an isomorphism
\begin{equation*}
    \mathrm{HH}^*_{\rm gr}(A, M)\cong{\rm Ext}^*_{A\mbox{-}{\rm gr}} (\K, {^{ad}(M)}),
\end{equation*}
where $\K$ is viewed as the trivial graded $A$-module via the
counit of $A$, $^{ad} M$ is the adjoint $A$-module associated to
the graded $A$-bimodule $M$ and $\mathrm{HH}^*_{\rm gr}$ denotes
the $G$-graded Hochschild cohomology. As an application, we deduce
that the graded cohomology of color Lie algebra $L$ is
isomorphic to the graded Hochschild cohomology of  its universal
enveloping algebra $U(L)$, solving
a question of M. Scheunert.
\end{abstract}

\section{Introduction}
Color Lie algebras have been introduced in \cite{RW} and studied
systematically in \cite{S1,S2,S3,SZ}. Some  recent interest relates
to
 their representation theory and related graded ring theory, \cite{CSV}.
 The Cartan-Eilenberg cohomology theory
  for Lie algebras \cite{Ch,CE}, has been extended to color Lie algebras by Scheunert and Zhang
  in \cite{S3,SZ}.
  In this note we introduce a graded cohomology of Color Lie algebras which coincide, in the case of degree zero, with the graded cohomology of $L$ defined by Scheunert and
Zhang. We show an isomorphism between the graded Hochschild cohomology of
  the universal enveloping algebra and the Lie cohomology for arbitrary color Lie
  algebras,
   which amounts to a careful manipulation of the group grading structure involved.
   We start from an abelian group $G$ with a skew-symmetric bicharacter
   $\varepsilon$. Consider a $G$-graded $\varepsilon$-Lie algebra
   $L$. Denote by $U(L)$ its  universal enveloping algebra.
   Note that $U(L)$ has a natural $(G,\varepsilon)$-Hopf algebra structure in the sense
      of Definition \ref{d2}. Therefore, first,  we study
       the cohomology of arbitrary $(G,\chi)$-Hopf algebras, where $\chi$ is any bicharacter of
       $G$. We prove the  main result concerning
       the (graded) Hochschild cohomology of a $(G,\chi)$-Hopf algebra
with bijective antipode (see Theorem \ref{t4}). 
 Then, we combine this with the information coming from the color
 Kozsul resolution of the trivial module $\K$
 of color Lie algebra
$L$ to get  the desired result for color Lie algebras (see Theorem
\ref{t5}). Our theorem extends the result of Cartan-Eilenberg for
Lie algebras (see \cite{CE}, pp. 277) and solves a question of M.
Scheunert in the case (of degree zero) (\cite{S3}). \par \vskip 5pt

This note is organized as follows: in Section 2 we fix notation
and provide background material concerning finite group gradings
and color Lie algebras; in Section 3 we study $(G,\chi)$-Hopf
algebras in detail and prove the main theorem Theorem \ref{t4}; in
Section 4 we study the color Koszul resolution of the trivial
module $\K$ of color Lie algebra $L$ (Theorem \ref{t3}), and, by
using it, we obtain Theorem \ref{t5}.\par \vskip 5pt

 We would like to thank Prof. Manfred
Scheunert very much to read this paper and give us many
suggestions. We also thank the referee for  helpful comments.

\section{Premilinaries}
Throughout this paper  groups are assumed to be abelian and $\K$ is
a field of characteristic zero. We recall some notation for graded
algebras and graded modules \cite{NV}, and some facts on color Lie
algebras from  \cite{S1,S2,S3,SZ}.

\subsection{Graded Hochschild cohomology}
Let $G$ be an abelian group with identity element $e$. We will
write $G$ as an multiplicative group.\par \vskip 5pt

An associative algebra $A$ with unit $1_A$, is said to be
$G$-graded, if there is a family $\left\{A_g|  g\in G\right\}$ of
subspaces  of $A$ such that $A=\oplus_{g \in G} A_g$ with  $1_A
\in A_e$ and $A_gA_h \subseteq A_{gh}$, for all $g, h \in G$. Any
element  $a \in A_g$ is  called homogeneous of degree $g$,
 and we write $|a|=g$. \par \vskip 5pt

A (left) graded $A$-module $M$ is a left $A$-module with an
decomposition $M=\oplus_{g \in G}M_g$ such that $A_g.M_h \subseteq
M_{gh}$. Let $M$ and $N$ be graded $A$-modules. Define
\begin{equation}
    \mathrm{Hom}_{A\mbox{-}{\rm gr}}(M,N)=
    \left\{f\in\mathrm{Hom}_{A}(M,N)|\ \  f(M_g)\subseteq N_g, \quad \forall\ \ g\in G\right\}.
\end{equation}
We obtain the category of graded left $A$-modules, denoted by
$A$-gr (see \cite{NV}). Denote by ${\rm Ext}_{A\mbox{-} {\rm
gr}}^n(-,-)$ the $n$-th right derived functor of the functor ${\rm
Hom}_{A\mbox{-} {\rm gr}}(-,-)$.\par \vskip 5pt

Let us recall the notion of graded Hochschild cohomology of a graded algebra $A$.
A graded $A$-bimodule is a $A$-bimodule $M=\oplus_{g \in G}M_g$ such that
 $A_g.M_h.A_k \subseteq M_{ghk}$. Similar as the above, we obtain the category of graded $A$-bimodules,
 denoted by $A\mbox{-}A \mbox{-}{\rm gr}$.\par \vskip 5pt

Let $A^e=A\otimes_\K A^{op}$ be the enveloping algebra of $A$,
where $A^{op}$ is the opposite algebra of $A$. Note that the
algebra $A^e$ also is graded by $G$ by  setting $A^e_g:=\sum_{h
\in G} A_h \otimes_\K A_{h^{-1}g}$.\par \vskip 5pt

 Now the graded $A$-bimodule $M$ becomes a graded
left $A^e$-module just by defining the $A^e$-action as
\begin{equation}
    (a \otimes a')m= a.m.a',
\end{equation}
and it is clear that $A^e_g  M_h \subseteq M_{gh}$, i.e., $M$ is a
graded $A^e$-module. Moreover, every graded left $A^e$-module arises
in this way. Precisely, the above correspondence establishes an
equivalence of categories
\begin{align}
    A\mbox{-}A \mbox{-} {\rm gr} \simeq A^e\mbox{-} {\rm gr}.
\end{align}
In the sequel we will identify these categories. \par \vskip 5pt

Let $M$ be a graded $A$-bimodule, equivalently, graded left
$A^e$-module. The $n$-th graded Hochschild cohomology of $A$ with
value in $M$ is defined by
\begin{align}
\mathrm{HH}_{\rm gr}^n (A, M):= {\rm Ext}_{A^e\mbox{-} {\rm gr}}^n
(A, M), \quad n \geq 0,
\end{align}
where $A$ is the  graded left $A^e$-module induced by the
multiplication of $A$, and the algebra $A^e=\oplus_{g \in G}
A^e_g$ is considered as a $G$-graded algebra as above.

\subsection{Color Lie algebras}
The concept of color Lie algebras is related to an abelian group $G$
and an anti-symmetric bicharacter $\varepsilon:G \times G
\rightarrow\K^\times$, i.e.,
 \begin{align}
&\varepsilon\left(g,h\right)\varepsilon\left(h,g\right)=1,\\
    &\varepsilon
\left(g,hk\right)=\varepsilon\left(g,h\right)\varepsilon\left(g,k\right),\\
    &\varepsilon\left(gh,k\right)=\varepsilon\left(g,k\right)\varepsilon\left(h,k\right),
\end{align}
where $g, h, k \in G$ and $\K^\times $ is the multiplicative group
of the units in $\K$.\par \vskip 5pt

 A $G$-graded space $L=\oplus_{g\in G} L_g$ is
said to be a $G$-graded $\varepsilon$-Lie algebra (or simply, color
Lie algebra), if it is endowed with a bilinear bracket
$\left[-,-\right]$ satisfying the following conditions
\begin{equation}
\left[ L_g, L_h \right]\subseteq L_{gh},
\end{equation}
\begin{equation}
    \left[a,b\right]=-\varepsilon\left(|a|,|b|\right)\left[b,a\right],
\end{equation}
\begin{equation}
\varepsilon\left(|c|,
|a|\right)\left[a,\left[b,c\right]\right]+\varepsilon\left(|a|,|b|\right)\left[b,\left[c,a\right]\right]+
\varepsilon\left(|b|, |c|\right)\left[c,\left[a,b\right]\right]=0,
\end{equation}
where $g, h \in G$,  and $a, b, c \in L$ are homogeneous
elements.\par \vskip 5pt

 For example, a
super Lie algebra is exactly  a $\Z_2$-graded $\varepsilon$-Lie
algebra where
\begin{equation}
    \varepsilon(i,j)=(-1)^{ij},\quad \forall\quad i,j\in\Z_2.
\end{equation}

Let $L$ be a color Lie algebra as above and $T(L)$ the tensor
algebra of the underlying $G$-graded vector space $L$. It is
well-known that $T\left(L\right)$ has a natural $\Z\times
G$-grading which is fixed by the condition that the degree of a
tensor $a_1\otimes...\otimes a_n$ with $a_i\in L_{g_i}$, $g_i\in
G$, $1\leq i\leq n$, is equal to $\left(n,g_1\cdots g_n\right)$.
The subspace of $T\left(L\right)$ spanned by homogeneous tensors
of order $\leq n$ will be denoted by $T^n\left(L\right)$. Let
$J\left(L\right)$ be the $G$-graded two-sided ideal of
$T\left(L\right)$ which is generated by
\begin{equation}\label{e1.7}
    a\otimes b-\varepsilon\left(|a|, |b|\right)b\otimes a-\left[a,b\right]
\end{equation}
with homogeneous $a, b \in L$. The quotient algebra
$U\left(L\right):=T\left(L\right)/J\left(L\right)$ is called the
universal enveloping algebra of the color Lie algebra $L$. The
$\K$-algebra $U\left(L\right)$ is a $G$-graded algebra and  has a
positive filtration by putting $U^n\left(L\right)$ equal to the
canonical image of $T^n\left(L\right)$ in $U\left(L\right)$.\par
\vskip 5pt
 In particular, if
$L$ is $\varepsilon$-commutative (i.e., $[L, L]=0$), then
$U\left(L\right)=S\left(L\right)$ (the $\varepsilon$-symmetric
algebra of the graded  space $L$). \par \vskip 5pt

The canonical map $\verb"i":L\rightarrow U\left(L\right)$ is a
$G$-graded homomorphism and satisfies
\begin{equation}
\verb"i" \left(a\right)\verb"i" \left(b\right)-\varepsilon\left(|a|,
|b|\right)\verb"i"\left(b\right)\verb"i"\left(a\right)=\verb"i"\left(\left[a,b\right]\right).
\end{equation}
The $\Z$-graded algebra $G(L)$ associated with the filtered
algebra $U\left(L\right)$ is defined by letting
$G^n\left(L\right)$ be the vector space
$U^n\left(L\right)/U^{n-1}\left(L\right)$ and $G\left(L\right)$
the space $\oplus_{n\in\N}G^n\left(L\right)$ (note
$U^{-1}\left(L\right):=\left\{0\right\}$). Consequently,
$G\left(L\right)$ is a $\Z\times G$-graded algebra. The well-known
generalized Poincar\'{e}-Birkhoff-Witt theorem, \cite{S1}, states
that the canonical homomorphism $\verb"i":L\rightarrow
U\left(L\right)$ is an injective $G$-graded homomorphism;
moreover, if $\left\{x_i\right\}_I$ is a homogeneous basis of $L$,
where the index set $I$ well-ordered. Set
$y_{k_j}:=\verb"i"\left(x_{k_j}\right)$, then the set of ordered
monomials $y_{k_1}\cdots y_{k_n}$ is a basis of $U\left(L\right)$,
where $k_j\leq k_{j+1}$ and $k_j< k_{j+1}$ if
$\varepsilon\left(g_j,g_j\right)\neq 1$ with $x_{k_j}\in L_{g_j}$
for all $1\leq j\leq n,n\in\N$. In case $L$ is finite-dimensional
$U\left(L\right)$ is a two-sided (graded) Noetherian algebra
(e.g., see \cite{CSV}).

\section{$(G,\chi)$-Hopf algebras, graded Hochschild cohomology}
Through this section $G$ is an abelian group with a bicharecter
$\chi: G \times G \longrightarrow \K^\times$. All unspecified graded
spaces (algebras, coalgebras, ...) are graded by $G$; all unadorned
${\rm Hom}$  and tensor  are taken over $\K$.

\subsection{Twisted algebras}

Let $(A=\oplus_{g\in G} A_g,\cdot, 1_A)$ be a graded algebra and
$\chi$ be a bicharacter. Then there exists a new (graded)
associative multiplication $\cdot^\chi$ on the  $\K$-space
$\oplus_{g\in G}A_g$ defined by
\begin{equation}
    a\cdot^\chi b=\chi(|a|, |b|)a\cdot b
\end{equation}
with $a,b$ homogeneous elements. It is easy to see that
$(\oplus_{g\in G}A_g,\cdot_\chi,1_A)$ is a (graded) associative
algebra, which will be  called the twisted algebra of $A$ by the
bicharacter $\chi$ and will be  denoted by $A^\chi$.\par \vskip 5pt

Let $A=(\oplus_gA_g,\cdot, 1_A)$ be a graded algebra and $\chi$ be
a bicharacter. Let $A^\chi$ be the  twisted algebra of $A$ by
$\chi$. Consider the opposite algebra $A^{op}$, and denote its
multiplication by $\circ$.  Thus we may consider the algebra
$(A^{op})^\chi$, the multiplication of which will be denoted by
$\cdot_\chi$. Hence we have
\begin{equation}\label{E2.2}
    a\cdot_\chi b=\chi(|a|, |b|) a \circ b=\chi(|a|, |b|) b\cdot a.
\end{equation}

Let $A=(\oplus_{g\in G}A_g,\cdot,1_A)$ be a graded algebra, and
$\chi$ a bicharacter and $A^\chi=(\oplus_{g\in G}A_g,\cdot^\chi,
1_A)$ the corresponding twisted algebra of $A$. Let
$M=\oplus_{g\in G}M_g$ be a graded $A$-module. Then there exists a
new graded $A^{\chi}$-module structure, denoted by $.^\chi$, on
the graded $\oplus_{g\in G} M_g$ defined by
\begin{equation}
    a.^\chi m :=\chi(|a|,|m|)a.m,
\end{equation}
where $a\in A$ and $m \in M$ are homogeneous. Thus $\oplus_{g\in
G} M_g$ becomes a right graded $A^{\chi}$-module, which will be
denoted by $M^\chi$. Clearly every  graded $A^{\chi}$-module
arises in the way. \par \vskip 5pt

Therefore we have

\begin{prop}
Use the above notation. There exists an equivalence of categories
between  $A$-{\rm gr} and ${A}^{\chi}$-{\rm gr}.
\end{prop}
\par
\vskip 5pt

 For further use, we need to introduce: let $A$ and $B$
be graded algebras (by $G$),   define a (graded) associative
algebra structure $(A\otimes B)^\chi$ on the space $A\otimes B$,
with the multiplication ``$*$'' given by the Lusztig's rule
\cite{L},
\begin{align}
(a \otimes b) * (a'\otimes b')= \chi(|b|, |a'|) aa' \otimes bb',
\end{align}
where $a, a' \in A$ and $b, b' \in B$ are  homogeneous.

\subsection{Twisted coalgebras}
 Recall from \cite{Sw} that a graded coalgebra $C$ is a graded space $C=\oplus_{g \in G} C_g$
with comultiplication  $\Delta:C\rightarrow C\otimes C$,  and counit
$\epsilon:C\rightarrow\K$ satisfying the following conditions: $
\Delta(C_g) \subseteq \sum_{h \in G} C_h \otimes C_{h^{-1}g}$,
 and $\epsilon(C_g)=0$ for $g\neq e$, $g \in G$.
 \par \vskip 5pt

We define twisted coalgebras as follows: let
$C=(C,\Delta,\epsilon)$ be a graded coalgebra, consider a new
(graded) comultiplication $\Delta^\chi$ on $C$ defined by
\begin{equation}
    ^\chi \Delta(c)=\sum\chi(|c_1|,|c_2|)(c_1\otimes c_2)
\end{equation}
where  $\Delta(c)=\sum_c {c_{1}} \otimes c_{2}$ is Sweedler's
notation with all factors $c_1$, $c_2$ homogeneous. It is easy
to check that $(C, { ^\chi \Delta},\epsilon)$ is a (graded)
colagebra, it will be denoted by $^\chi C$.\par \vskip 5pt

Note that the opposite coalgebra of $^\chi C$, denoted by $(^\chi
C)^{cop}$ will have the comultiplication as follows
\begin{equation}
    _\chi \Delta(c)=\sum\chi(|c_1|,|c_2|)(c_2\otimes c_1).
\end{equation}

Let $C$ be a graded coalgebra. Denote by {\rm gr}-$C$ the category
of graded right $C$-comodules, with morphisms being graded
homomorphism of comodules (of degree $e$).\par \vskip 5pt

 Dually to
Proposition 1, we obtain

\begin{prop}Use the above notation. There exists an equivalence of categories
 between {\rm gr-}$C$ and {\rm gr-}$^\chi C$.
\end{prop}

\par \vskip 5pt
We need the following construction: let $(C, \Delta_C,
\epsilon_C)$ and $(D, \Delta_D, \epsilon_D)$ be two graded
coalgebras, then the following law
\begin{align}
^\chi (\Delta_C\otimes \Delta_D)(c \otimes d)= \sum \chi(|c_{2}|,
|d_{1}|) \hskip 2pt (c_{1} \otimes d_{1}) \otimes (c_{2} \otimes
d_{2})
\end{align}
defines on the space $C\otimes D$ a structure of graded coalgebra
with counit $\epsilon_C \otimes \epsilon_D$, which is denoted  by
$^\chi (C \otimes D)$.

\subsection{$(G,\chi)$-Hopf-algebras}

\begin{de}\label{d2} A $(G, \chi)$-Hopf algebra $A$ (compare \cite{LZ} and \cite{M}, p.206)  is a 5-tuple $(A, m, \eta, \Delta, \epsilon, S)$ such that  \\
(\textbf{T}1): $A=\oplus_{g \in G} A_g$ is a graded algebra with
multiplication $m : A \otimes A \longrightarrow A$ and
 the unit map $\eta: K \longrightarrow A$. In the meantime, $(A, \Delta, \epsilon)$ is a graded coalgebra with
 respect to the same grading.\\
(\textbf{T}2): The counit $\epsilon: A \longrightarrow K$ is an algebra map.
The comultiplication $\Delta: A \longrightarrow (A \otimes A)^\chi$ is an algebra map,
where the algebra $(A \otimes A)^\chi$ is defined as in (3.4). \\
(\textbf{T}3): The antipode  $S: A\longrightarrow A$ is a graded map
such that
 \begin{align}\label{e2.1}
 \sum a_{1} S(a_{2})=\epsilon(a)=\sum S(a_{1})a_{2}
 \end{align}
 for all homogeneous $a \in A$, where we use Sweedler's notation $\Delta(a)=\sum a_{1}\otimes a_{2}$.
\end{de}

\begin{rem}\label{n21}\
\begin{enumerate}
    \item A 4-tuple $(A, m, \eta, \Delta, \epsilon)$ satisfying the (\textbf{T}1) and (\textbf{T}2) will be called a $(G, \chi)$-bialgebra.
    \item The condition (\textbf{T}2) implies exactly that the following holds:
\begin{align}
 &\epsilon(1_A)=1, \quad \epsilon(aa')=\epsilon(a) \epsilon(a'), \\
 &\Delta(1_A)=1_A \otimes 1_A,\\
 &\Delta(aa')= \sum \chi(|a_{2}|, |a'_{1}|)\hskip 2pta_{1} a'_{1}\otimes a_{2}a'_{2}=\Delta(a)*\Delta(b),
 \end{align}
where $1_A$ is the identity element of $A$, and $a, a' \in A$ are
homogeneous. Note that these four equations exactly state that
\begin{equation}
    \eta:\K \longrightarrow A, \quad m:{^\chi (A \otimes A)}\longrightarrow A
\end{equation}
are coalgebra maps, where the coalgeba $^\chi (A \otimes A)$ is
defined in (3.6).
  \item A Hopf ideal of a $(G, \chi)$-Hopf algebra $A$ is a graded ideal
$I \subseteq A$ and a coideal (i.e., $\Delta(I) \subseteq A
\otimes I + I \otimes A$ and $\epsilon(I)=0$) satisfying $S(I)
\subseteq I$.\par \vskip 5pt

Thus there exist a unique $(G, \chi)$-Hopf algebra structure on the
space  $A/I$ such that the canonical map $\pi: A \longrightarrow
A/I$ is a $(G, \chi)$-Hopf algebra morphism.
 \end{enumerate}
 \end{rem}

\vskip 10pt

Let $(A, m, \eta)$ and $(C, \Delta, \epsilon)$ be a graded algebra and a graded coalgebra respectively.
 Then ${\rm Hom}(C, A)$ becomes an associative algebra with the  convolution product $\star$ defined by
 \begin{align}
 (f \star g)(c)&= m \circ (f \otimes g ) \circ \Delta (c)=\sum f(c_{1})g(c_{2}),
  \end{align}
  for $f, g \in {\rm Hom}(C, A)$, $c\in C$. Note that the unit of ${\rm Hom}(C,A)$
  is $\eta\circ\epsilon$. Moreover, it is easy to see that ${\rm Hom}_{\rm gr} (C, A)$
  is a subalgebra of ${\rm Hom}(C, A)$, i.e., if $f$ and $g$ are graded maps (of degree $e$), then so is $f\star
  g$.\par \vskip 5pt

Let $A=(A, m, \eta, \Delta, \epsilon, S)$ be a $(G, \chi)$-Hopf
algebra. Consider the algebra ${\rm Hom} (A, A)$ with the
convolution product $\star$. Then the condition (\textbf{T}3) is
equivalent to
\begin{align}
S \star {\rm Id_A}=\eta\circ\epsilon ={\rm Id_A} \star S.
\end{align}
This shows the uniqueness of the antipode $S$. Now we obtain a
result similar to the one in \cite{Sw}, p.74 (compare \cite{LZ},
Theorem 2.10).

\begin{lem}\label{n2.1}\
\begin{enumerate}
    \item The antipode $S: (A,\cdot, 1_A) \longrightarrow ((A^{op})^{\chi},\cdot_\chi, 1_A)$
    is an algebra morphism with $\cdot_\chi$ defined by equation (\ref{E2.2}).
  \item The antipode $ S: (^\chi A)^{cop} \longrightarrow A$ is a coalgebra
  morphism, where the comultiplication of $(^\chi A)^{cop}$ is defined by (3.6).
\end{enumerate}
\end{lem}

\begin{proof} We will imitate the proof in \cite{Sw}, and we will  only prove the first statement, since the second
 can be proved similarly. \par \vskip 5pt
 One see $S(1_A)=1_A$ by the condition (\textbf{T}3).  Now it suffices to show that
  $S(aa')= S(a)\cdot_\chi S(a')= \chi(|a|, |a'|) S(a')S(a)$ for all homogeneous elements $a, a' \in A$.
Consider the convolution algebra ${\rm Hom}({^\chi (A \otimes A)},
A)$, where the coalgebra structure of $^\chi (A\otimes A)$ is
defined as in (3.7). Now define two elements $F, G \in {\rm
Hom}({^\chi (A \otimes A)}, A)$ by
\begin{align}
F(a\otimes a')= S(aa') \quad \mbox{and} \quad G(a \otimes a')=
\chi(|a|, |a'|) S(a')S(a).
\end{align}
So we need to show that $F=G$.  We claim that
\begin{align}F \star m=m\star G= \eta\circ (\epsilon \otimes \epsilon)
\end{align}
in the convolution algebra ${\rm Hom}({^\chi (A \otimes A)}, A)$,
where $m$ denotes the multiplication of $A$. Indeed that $\eta
\circ (\epsilon\otimes \epsilon)$ is the unit in the convolution
algebra
 ${\rm Hom}({^\chi (A \otimes A)}, A)$, thus we obtain $F=G$, as
 required.\par \vskip 5pt
In fact, by the definition of the convolution product $\star$ and
the coalgebra structure of $^\chi (A\otimes A)$, we have
\begin{align}
(F \star m) (a \otimes a')&= m \circ (F \otimes m) \circ {^\chi (\Delta \otimes \Delta)} (a \otimes a')\\
                           &= \sum \chi(|a_{2}|, |a'_{1}|)\hskip 2pt  S(a_{1}a'_{1}) a_{2}a'_{2} \quad \mbox{By Remark \ref{n21} } \\
                           &=m (S\otimes {\rm Id}) \Delta(aa')\\
                           &=\epsilon(ab)= (\eta\circ (\epsilon \otimes \epsilon))(a\otimes a').
\end{align}
On the other hand,
\begin{align}
(m \star G) (a \otimes a') &= m \circ (m \otimes G)\circ {^\chi (\Delta \otimes \Delta)} (a \otimes a')\\
                            &= \sum \chi(|a_{2}|, |a'_{1}|) \hskip 2pt a_{1}a'_{1}  G(a_{2}\otimes a'_{2})\\
                          &=\sum  \chi(|a_{2}|, |a'_{1}|) \chi(|a_{2}|, |a'_{2}|) \hskip 2pt a_{1}a'_{1} S(a'_{2}) S(a_{2})\quad \mbox{By  } |a'|=|a'_{1}| |a'_{2}|\\
                          &=\sum \chi(|a_{2}|, |a'|)\hskip 2pt  a_{1}a'_{1} S(a'_{2}) S(a_{2})\\
                          &= \sum \chi (|a_{2}|, |a'|) a_{1}S(a_{2}) \epsilon(a')\\
                          &= \epsilon (a) \epsilon(a')=(\eta\circ (\epsilon \otimes \epsilon))(a\otimes a').
\end{align}
(Here the second to the last equality follows from the fact that
if  $|a'|\neq e$ then  $\epsilon(a')=0$; otherwise,
$\chi(|a_{2}|, |a'|)=1$.) Thus we arrive at
\begin{equation}
    (F \star m)=\eta \circ (\epsilon\otimes   \epsilon)=(m \star G).
\end{equation}
This completes the proof.
\end{proof}

\begin{rem}\label{r2.2}
If the antipode $S$ is bijective with inverse $S^{-1}$, then by
Lemma \ref{n2.1}, we have:
\begin{equation}
     S^{-1}(a) S^{-1}(a')=\chi(|a|, |a'|)S^{-1}(a'a).
\end{equation}
In this case, we call such a $(G,\chi)$-Hopf algebra a color Hopf
algebra.
\end{rem}
\par \vskip 5pt

The following result is quite useful when we construct an antipode on a $(G, \chi)$-bialgebra.
\begin{lem}\label{n2.2}
Let $(A, m, \eta, \Delta, \epsilon)$ be a $(G, \chi)$-bialgebra
generated by a set $\Lambda$ of homogeneous elements (as an
algebra). If there exists an algebra morphism $S: A
\longrightarrow (A^{op})^{\chi}$ such that each (\ref{e2.1}) holds
for each   $a \in \Lambda$, then $S$ is the antipode of $A$.
\end{lem}
\begin{proof} We just need to check the (\ref{e2.1}) for all elements in $A$. For this, it suffices to show that if two homogeneous elements $a, b \in A $
satisfy the  (\ref{e2.1}), so does $ab$.\par \vskip 5pt

In fact, by Remark \ref{n21} and then Lemma \ref{n2.1}
\begin{align}
\sum (ab)_1 S((ab)_2)&= \sum \chi(|a_2|, |b_1|)a_1b_1 S(a_2b_2)\\
                     &= \sum  \chi(|a_2|, |b_1|) \chi (|a_2|, |b_2|)a_1b_1 S(b_2) S(a_2)\\
                     &=\sum \chi(|a_2|, |b|) \epsilon(a)\epsilon(b)\\
                     &=\epsilon(a) \epsilon(b)=\epsilon(ab).
\end{align}
(The third  equality uses the fact that $|b|=|b_1|\cdot |b_2|$; the
fourth equality uses the fact that $\epsilon(b)\neq 0$ implies
$\chi(|a_2|, |b|)=1$.) In a similar way we may establish the right
hand side of the  (\ref{e2.1}). This completes the proof.
\end{proof}
\par \vskip 5pt

\subsubsection*{Example}\label{E}
Using Lemma \ref{n2.2}, we will give an important example of color
Hopf algebras. Let $V$ be a $G$-graded space and denote by $T(V)$
the tensor algebra on $V$. Thus $T(V)$ is a $G$-graded algebra
generated by $V$. By the universal property of $T(V)$, there exist
unique graded algebra morphisms
\begin{equation}
    \Delta: TV \longrightarrow (T(V)\otimes T(V))^\chi\quad v\longmapsto 1\otimes v + v\otimes 1,
\end{equation}
\begin{equation}
    \quad \epsilon : T(V) \longrightarrow \K\quad  v\longmapsto 0,
\end{equation}
for all $ v \in V$. \par \vskip 5pt To show that $(T(V), \Delta,
\epsilon)$ is a (graded) coalgebra, we need to verify
\begin{align}
(\Delta \otimes {\rm Id})\circ\Delta=({\rm Id} \otimes
\Delta)\circ \Delta \quad \mbox{and} \quad (\epsilon \otimes {\rm
Id})\circ\Delta={\rm Id}= ({\rm Id} \otimes \epsilon)\circ\Delta
\end{align}
where ${\rm Id}$ the identity map of $T(V)$. \par \vskip 5pt

Note that all above maps are algebra morphisms, so it suffices to
check them on a set of generators of the algebra $T(V)$. Clearly
all elements in $V$ satisfy the above equations, thus we have
shown that $(T(V), \Delta, \epsilon)$ is an coalgebra, hence $TV$
is a $(G, \chi)$-bialgebra. \par \vskip 5pt Again by the universal
property of $T(V)$, there exists a unique algebra map
\begin{equation}
    S: T(V) \longrightarrow (T(V)^{op})^{ \chi} \quad v\longmapsto-v,
\end{equation}
for all $v \in V$. Now applying Lemma \ref{n2.2}, we deduce that
$T(V)$ is a color Hopf algebra. We call the resulting color Hopf
algebra $T(V)$ the tensor color Hopf algebra of $V$.\par \vskip
5pt
 It follows from Remark 1
that if $I$ is a Hopf ideal of $T(V)$, then we have a quotient
$(G, \chi)$-Hopf algebra $T(V)/I$.\par \vskip 5pt

 An important example is as
follows: let $L$ be a $G$-graded $\varepsilon$-Lie algebra, then
its universal enveloping algebra $U(L)=T(L)/J(L)$ is $(G,
\varepsilon)$-Hopf algebra with $J(L)$ defined by (\ref{e1.7}),
since $J(L)$ is a Hopf idea. Explicitly, $U(L)$ is a color Hopf
algebra with comultiplication $\Delta$ and $\epsilon$ given by
\begin{align}\label{e36}
\Delta(a)= 1 \otimes a+ a \otimes 1, \quad \epsilon(a)=0, \quad
\forall\quad  a \in L.
\end{align}\par
\vskip 10pt

We now consider graded right $A$-modules. As before $A$ will be a
$(G, \chi)$-Hopf algebra. Recall that a right gr-free $A$-module
of the form $V \otimes A$, where $V$ is a graded space and $V
\otimes A$ is graded by assigning to $v \otimes a$ the degree $|v|
\cdot |a|$, for all homogeneous elements $v \in V$ and $a \in A$,
and the right action is given  by (see \cite{NV})
\begin{equation}
    (v \otimes a) a'= v \otimes aa'.
\end{equation}
 In fact, gr-free modules are just the free objects in the category of graded right
 $A$-modules.\par \vskip 5pt

Since  $\Delta: A \longrightarrow (A \otimes A)^\chi$ is an algebra map, the
algebra $(A \otimes A)^\chi$ becomes a graded right $A$-module.
Explicitly, the right $A$-action on $(A\otimes A)^\chi$ is given by

\begin{align}
(a\otimes a')b:= (a\otimes a')*\Delta(b)=\sum \chi(|a'|, |b_{1}|) a b_{1} \otimes a' b_{2},
\end{align}
for homogeneous $a, a', b \in A$ and $\Delta(b)=\sum b_1 \otimes
b_2$ is the Sweedler notation. Note that the  grading of
$(A\otimes A)^\chi$ is given such that the degree of $a\otimes a'$
is $|a|  \cdot |a'|$. \par\vskip 5pt

 The following result will be
essential.\par
\begin{prop}\label{p2.1}
The right $A$-module  $(A\otimes A)^\chi$ defined above  is gr-free.
\end{prop}

\begin{proof} Let $V$ denote the underlying graded space of $A$. Thus $V \otimes A$ becomes a right gr-free module. Define a map  $\Psi: (A\otimes A)^\chi \longrightarrow V \otimes A$ by
\begin{align}
\Psi(a \otimes a')= \sum a S(a'_{1})\otimes a'_{2},
\end{align}
where $a, a' \in A$ are homogeneous and $\Delta(a)=\sum a'_1
\otimes a'_2$ is the Sweedler notation. It is obvious that $\Psi$
is a  (graded) bijective map with inverse
$$\Psi^{-1}(a \otimes a')= \sum aa'_{1}\otimes a'_{2}.$$
We claim that $\Psi$ is a right $A$-module morphism, then we are
done.\par \vskip 5pt

 In fact we have
\begin{align}
\Psi((a\otimes a')b)&= \sum  \chi(|a'|, |b_{1}|) \Psi(ab_{1}\otimes a'b_{2})\\
                   &=\sum \chi(|a'|, |b_{1}|) ab_{1}S((a'b_{2})_{1}) \otimes (a'b_{2})_{2}.
\end{align}
By Remark \ref{n21} we have
\begin{align}
\Psi((a \otimes a')b)=\sum \chi(|a'|, |b_{1}|) \chi(|a'_{2}|, |b_{2}|) ab_{1} S(a'_{1}b_{2}) \otimes a'_{2}b_{3},
\end{align}
where we use $(\Delta \otimes {\rm Id}_A)\Delta(b)=\sum b_1
\otimes b_2 \otimes b_3$ (see \cite{Sw}). \par

 By Lemma \ref{n2.1}, we have
$S(a'_{1}b_{2})= \chi (|a'_{1}|, |b_{2}|) S(b_{2})S(a'_{1})$,
hence
\begin{align}
\Psi((a\otimes a')b)&= \sum \chi(|a'|, |b_{1}|) \chi (|a'_{2}|, |b_{2}|) \chi (a'_{1}, b_{2})ab_{1}S(b_{2}) S(a'_{1}) \otimes a'_{2}b_{3}\\
                   &=\sum \chi (|a'|, |b_{1}||b_{2}|) ab_{1}S(b_{2}) S(a'_{1}) \otimes a'_{2}b_{3}\\
                   &=\sum \chi (|a'|, |b_{1}|) a \epsilon (b_{1}) S(a_{1}) \otimes a'_{2}b_{2}.
\end{align}
Using the fact that $\epsilon(b_{1})\neq 0$ implies that  $|b_{1}|=e$ and hence $\chi(|a'|, |b_{1}|)=1$, we obtain
\begin{align}
\sum \chi (|a'|, |b_{1}|) a \epsilon (b_{1}) S(a_{1}) \otimes a'_{2}b_{2}&= \sum aS(a'_{1})\otimes a'_{2}b_{2}\\
                           &=\sum a S(a'_{1})\otimes a'_{2}b.
\end{align}
 Note the right $A$-module structure on $V \otimes A$. So we have proved that
\begin{align}
\Psi((a\otimes a')b)= (\Psi(a \otimes a'))b,  \quad \forall \quad a, a', b \in A.
\end{align}
This completes the proof.
\end{proof}
\vskip 10pt

 We obtain
 \begin{thm}
Let $(A, m,\eta, \Delta, \epsilon, S)$ be a color Hopf algebra.
Then the categories $A \mbox{-} A \mbox{-}{\rm gr}$ and $(A\otimes
A)^\chi \mbox{-}{\rm gr}$ are equivalent.
 \end{thm}
\begin{proof}We are going to construct the functor
 \begin{align}
 F: A \mbox{-} A \mbox{-gr} \rightsquigarrow (A\otimes A)^\chi \mbox{-gr}
 \end{align}
as follows: let $M$ be a graded $A$-bimodule, we denote the two-sided $A$-action on $M$ by ``.''. Define $F(M)=M$ as graded spaces with the left $(A \otimes A)^\chi$-action  given by
 \begin{align}
  (a\otimes a') m= \chi (|a'|, |m|) a.m. S(a'),
 \end{align}
 where $a, a' \in A$ and $m \in M$ are homogeneous.
 We claim that the action is well-defined, i.e.,
 \begin{align}
 ((a\otimes a') * (b \otimes b'))m= (a\otimes a') ((b \otimes b')m),
 \end{align}
 where $*$ denotes the multiplication of the algebra $(A \otimes A)^\chi$ (see (3.4)).
 \par \vskip 5pt

 In fact, we have
 \begin{align}
  ((a\otimes a') *(b \otimes b'))m &= \chi (|a'|, |b|) (ab \otimes a'b')m\\
                                        &= \chi (|a'|, |b|) \chi (|a'b'|, |m|) ab. m .S(a'b')\\
                                        &= \chi (|a'|, |b|) \chi (|a'b'|, |m|) \chi (|a'|, |b'|) ab.m. S(b')S(a').
 \end{align}
The last equality uses Lemma \ref{n2.1}. 1. On the other hand,
 \begin{align}
  (a\otimes a') ((b \otimes b')m) &= \chi (|b'|, |m|) (a\otimes a') b.m.S(b')\\\
                                 &=\chi (|b'|, |m|) \chi (|a'|, |b.m.S(b')|) a.(b.m. S(b')).S(a').
 \end{align}
Note that the degree of the element $b.m.S(b')$ is $|b|\cdot
|m|\cdot |b'|$. By comparing the above two identities, we have
proved the claim. \par \vskip 3pt

Conversely, we have  the functor
 \begin{align}
 G: (A\otimes A)^\chi \mbox{-gr}  \rightsquigarrow  A \mbox{-}A \mbox{-gr}
 \end{align}
 given as follows: let $N$ be a left graded $(A\otimes A)^\chi$-module, define $G(N)$ to be
 $N$ as graded spaces, and its $A$-bimodule structure given by
 \begin{align}
 a.n =(a\otimes 1) n \quad  \mbox{ and } \quad a.a'=\chi^{-1}(|a'|, |n|) (1\otimes S^{-1}(a')) n,
 \end{align}
 for all homogeneous $a , a' \in A$ and $n \in N$. Clearly, $G(N)$ is a left $A$-module. Note that
 \begin{align}
 (n.a).b&= \chi^{-1} (|b|, |a|\cdot |n|) (1\otimes S^{-1}(b))(n.a)\\
        &=\chi^{-1} (|b|, |a| \cdot |n|)\chi ^{-1} (|a|, |n|) ((1\otimes S^{-1}(b))\star (1\otimes S^{-1}(a))) n\\
         &= \chi^{-1} (|b|, |a|\cdot |n|)\chi ^{-1} (|a|, |n|) \chi (|b|, |a|) (1\otimes S^{-1}(ab))n\\
         &=\chi^{-1}(|ab|, |n|) (1 \otimes S^{-1}(ab))n = n.(ab).
 \end{align}
 The third equality uses the fact $S^{-1}(b) S^{-1}(a)= \chi (|b|, |a|) S^{-1}(ab)$, see Remark \ref{r2.2}, hence  $G(N)$ is also a right $A$-module. Note that  $(a.n).a'=a.(n.a')$, therefore, $G(N)$ is a graded $A$-bimodule.
It is easy to check that the functors $F$ and $G$ are inverse to each other. Thus we have proved the result.
\end{proof}

\subsection{Twisted Tensor Modules}
In this subsection, we include some remarks and notation
concerning tensor modules in the category $A$-gr for a given $(G,
\chi)$-bialgebra $A$. \par \vskip 5pt

 For given graded
$A$-modules $M=\oplus_{g \in G} M_g$ and $N=\oplus_{g \in G} N_g$,
we define a graded $A$-module $(M \otimes^\chi N)$  as follows: as
graded spaces $(M \otimes^\chi N)$ coincides with $M\otimes N$
(note that the degree of $m \otimes n$ is just $|m|\cdot |n|$ for
homogeneous $ m\in M$ and $n \in N$); the (left) $A$-action is
given by
\begin{align}
a (m \otimes n):= \sum \chi (|a_2|, |m|)a_1 m \otimes a_2n,
\end{align}
for all homogeneous elements $a \in A, m \in M, n \in N$ and
$\Delta(a)=\sum a_1 \otimes a_2$ is the Sweedler notation.  It is
easy to check that $(M \otimes^{\chi} N)$ is a graded
$A$-module.\par \vskip 5pt

 Since $\epsilon : A \longrightarrow \K$ is an algebra
map, then $\K$ becomes a  graded $A$-module, which will be
referred to as the trivial module (note that $\K$ is  trivially
graded). Thus the above defined ``$(\otimes)^{\chi}$'' has the
following properties:
 \begin{align*}
 ((M
\otimes^{\chi} N) &\otimes^{\chi}  L)\simeq (M \otimes^{\chi} (N \otimes^{\chi} L)),\\
 (M \otimes^{\chi}  \K) & \simeq M \simeq (\K \otimes^{\chi}  M),
 \end{align*}
 for all $M, N, L \in $ $A$-gr.

\subsection{Graded Hochschild Cohomology}
 Throughout this subsection, assume that $(A, m, \eta, \delta, \epsilon, S)$ is a color Hopf algebra.\par \vskip 5pt

  Denote by $A$-gr (resp. $(A \otimes A)^\chi$-gr) the category of
 (left) graded $A$-modules (rep. $(A \otimes A)^\chi$-modules) with graded morphisms (of  degree $e$).
 Note that here $(A\otimes A)^\chi$ is considered as an $G$-graded algebra as above. \par \vskip 5pt

 Since $\Delta: A \longrightarrow (A\otimes A)^\chi$ is an algebra map, there is a restriction  functor
 \begin{align}
 {\rm Res}: (A \otimes A)^\chi{\rm  \mbox{-}gr} \rightsquigarrow A{\rm  \mbox{-}gr}.
 \end{align}
More precisely, if $M$ is a graded $(A \otimes A)^\chi$-module,
 then ${\rm Res} (M)=M$ as graded spaces, and its left $A$-action is given by
\begin{align}\label{e2.3}
a m= (\sum a_1 \otimes a_2)m, \quad a \in A, \quad m \in M.
\end{align}
\vskip 5pt

The following is a direct consequence of Proposition \ref{p2.1}.
\begin{prop}
The functor ${\rm Res}: (A \otimes A)^\chi{\rm  \mbox{-}gr} \rightsquigarrow A{\rm  \mbox{-}gr}$ is exact and it preserves injective objects.
\end{prop}
\begin{proof} The exactness of ${\rm Res}$ is obvious. Let $I$ be an injective object in $(A \otimes A)^\chi{\rm  \mbox{-}gr}$. To show that ${\rm Res}(I)$ is an injective object in $A {\rm \mbox{-}gr}$,  take any monomorphism
\begin{equation}
    i: N \longrightarrow N'
\end{equation}
 and a morphism
\begin{equation}
    f: N \longrightarrow  {\rm Res}(I)
\end{equation}
in $A {\rm \mbox{-}gr}$, we claim that there exists some morphism
\begin{equation}
    f' : N' \longrightarrow {\rm Res}(I)
\end{equation}
such that $f'i=f$. then we are done. In fact, by Proposition \ref{p2.1}, $(A \otimes A)^\chi$ is a gr-free right $A$-module, hence it is flat $A$-module. So we have a monomorphism of left $(A\otimes A)^\chi$-modules
 \begin{align}
j=(A \otimes A)^\chi \otimes_A i: (A \otimes A)^\chi \otimes_A N \longrightarrow (A \otimes A)^\chi \otimes_A N'.
 \end{align}
Note that we have a morphism of right $(A\otimes A)^\chi$-modules
\begin{align}
g: (A \otimes A)^\chi \otimes_A N \longrightarrow I,\quad (a \otimes a') \otimes n\longmapsto (a \otimes a') f(n),
\end{align}
where $a, a' \in A$ and  $n \in N$. Since $I$ is an injective object in $(A \otimes A)^\chi{\rm  \mbox{-}gr}$, there exist a morphism
\begin{equation}
    g': (A \otimes A)^\chi \otimes_A N' \longrightarrow I\quad\mathrm{such\quad that}\quad g' \circ j=g.
\end{equation}
 Define
\begin{equation}
    f': N' \longrightarrow {\rm Res} (I),\quad n'\longmapsto g'((1_A\otimes 1_A)\otimes n'),
\end{equation}
 where $n' \in N$ and $1_A \in A$ is the unit. Now it is easy to check that $f' i=f$ and this proves that ${\rm res} (I)$ is an injective object in $A \mbox{-gr}$.
\end{proof}
\par \vskip 5pt

 Define the adjoint functor to be
$$^{ad }(-)={\rm Res} \circ F: A\mbox{-} A\mbox{-gr} \rightsquigarrow A\mbox{-gr}.$$
Explicitly, let $M$ be a graded $A$-bimodule,
 then $^{ad}(M)=M$ as graded spaces, and the left $A$-module structure is given by
\begin{align}\label{e2.5}
 am= \sum \chi (|a_{(2)}|, |m|) a_{1}.m. S(a_{2}),
\end{align}
for homogeneous $a\in A$ and $m \in M$. The resulting graded
$A$-module $^{ad} (M)$ is called the adjoint module associated the
graded $A$-bimodule $M$.\par \vskip 10pt

 The main theorem in this section is as follows:
\begin{thm}\label{t4}
Let $A=(A, m, \eta, \Delta, \epsilon, S)$ be a color Hopf algebra
and let $M$ be a graded $A$-bimodule.  Then there exists an
isomorphism
\begin{align*}
\mathrm{HH}_{\rm gr}^n(A, M)\simeq{\rm Ext}^n_{A \mbox{-} {\rm gr}}
(\K, {^{ad}(M)}),\quad n\geq 0,
\end{align*}
where $\K$ is viewed as the trivial graded $A$-modules via the
counit $\epsilon$, and $^{ad} (M)$ is the adjoint $A$-module
associated to the graded $A$-bimodule $M$.
\end{thm}

\begin{proof} First we show that there exists a natural
isomorphism
\begin{align*}
{\rm Hom}_{A^e\mbox{-}{\rm gr}} (A, -)\simeq {\rm
Hom}_{A\mbox{-}{\rm gr}} (\K, {^{ad}(-)}),
\end{align*}
both of which are functors form $A$-$A$-gr to the category of vector
spaces.\par \vskip 3pt

In fact, for each graded $A$-bimodule $M=\oplus_{g \in G}{M_g}$,
\begin{align*}
{\rm Hom}_{A^e\mbox{-}{\rm gr}} (A, M)=\{m \in M_e| a.m=m.a, \quad
\mbox{for all }a\in A\},
\end{align*}
and
\begin{align*}
{\rm Hom}_{A\mbox{-}{\rm gr}} (\K, {^{ad}(M)})= \{m \in M_e| am=
\epsilon(a)m, \quad \mbox{for all } a\in A\}.
\end{align*}
We deduce that the isomorphism in using  the definition of $am$,
see (\ref{e2.5})).\par \vskip 3pt

In general, for $n\geq 1$, we have
\begin{align*}
\mathrm{HH}_{\rm gr}^n(A, -)&={\rm Ext}^n_{A^e\mbox{-}{\rm gr}} (A,-)\\
&= R^n {\rm Hom}_{A^e\mbox{-}{\rm gr}}(A, -),
\end{align*}
where $R^n$ means taking the $n$-th right derived functors. Now
apply the above observation, we get

\begin{align*}
   \mathrm{HH}_{\rm gr}^n(A, -)=R^n ({\rm Hom}_{A\mbox{-}{\rm gr}} (\K, -)\circ
   {^{ad}(-)}).
   \end{align*}
By Theorem 1 and Proposition 4, we obtain the functor $^{ad}(-)$
is exact and preserves injective objectives. Hence Grothendieck's
spectral sequence (e.g., see \cite{CE} or [HS], p. 299) gives us

 \begin{align*}
R^n ({\rm Hom}_{A\mbox{-}{\rm gr}} (\K, -)\circ
   {^{ad}(-)})= R^n ({\rm Hom}_{A\mbox{-}{\rm gr}} (\K, -))\circ
   {^{ad}(-)}.
\end{align*}
Hence, we have

   \begin{align*}
   \mathrm{HH}_{\rm gr}^n(A, -) &=R^n {\rm Hom}_{A\mbox{-}{\rm gr}} (\K, -) \circ {^{ad} (-)} \\
                          &={\rm Ext}^n_{A\mbox{-}{\rm gr}}(\K, -) \circ {^{ad}(-)} \\
                          &={\rm Ext}^n_{A\mbox{-}{\rm gr}}(\K, {^{ad}(-)}).
\end{align*}
This completes the proof.
\end{proof}
\par

\vskip 10pt
\subsection{Shift functor}
We end this section with some observations on shift functor. Let
$A$ be a $G$-graded algebra. For each $h \in G$, we define a shift
functor $[h]$ from $A$-gr to itself as follows: for each $M \in
A$-gr, define a graded $A$-module $M[h]$ by setting
$(M[h])_g=M_{hg}$ for each $g \in G$. Note that $M[e]=M$.\par
\vskip 5pt

 Set
\begin{equation}
    \mathrm{HOM}_{A}(M,N)_h=\left\{f\in\mathrm{Hom}_{A}(M,N)
    |\ \  f(M_g)\subseteq N_{gh}, \forall\ \ g\in G\right\}.
\end{equation}
So we have (see \cite{NV}, pp. 25):
\begin{equation}\label{e7.5}
  \mathrm{HOM}_{A}(M,N)_h=\mathrm{Hom}_{A\mbox{-}{\rm gr}}(M,N[h]).
\end{equation}
 For $M, N \in A$-gr,
set
\begin{align}
{\rm HOM}_A (M, N):=\oplus_{g \in G} {\rm HOM}_{A}(M , N)_h.
\end{align}

Let ${\rm EXT}_A^n (-, -)$ (resp. ${\rm EXT}_A^n (-, -)_h$) be the
$n$-th right derived functor of the functor ${\rm HOM}_{A}(-, -)$
(resp. ${\rm HOM}_{A}(-, -)_h$).\par \vskip 5pt

 Clearly, we have
\begin{align}
{\rm EXT}_A^n (M, N)_h={\rm Ext}_{A\mbox{-gr}}^n(M , N[h]),\quad n
\geq 0
\end{align}
and, consequently,
\begin{align}\label{e12}
{\rm EXT}_A^n (M, N)=\oplus_{h \in G}\;  {\rm
Ext}_{A\mbox{-gr}}^n(M , N[h]),\quad n \geq 0.
\end{align}

Let $M$ be a graded $A$-bimodule which is regarded as a left
$A^{e}$-module. \par
Set
\begin{equation}
\mathrm{HH}^n (A, M)_h:= {\rm EXT}^n _{A^e}(A,
M)_h=\mathrm{HH}_{\rm gr}^n (A, M[h]),\quad  n \geq 0.
\end{equation}
\par \vskip 5pt

From Theorem \ref{t4}, we get immediately
\begin{cor}\label{c1} Under the hypotheses of Theorem \ref{t4}. Then
$$\mathrm{HH}^n (A, M)_h\simeq {\rm Ext}^n_{A \mbox{-} {\rm gr}}(\K, {^{ad}(M[h])})=\mathrm{EXT}^n_{A}(\K,^{ad}({M}))_h
$$
for each $h\in G$ and  $n\geq 0$.
\end{cor}

\par \vskip 5pt


\begin{rem}
In the above two corollaries, we use the notation $^{ad}(M[h])$.
Let us remark that, in general, $^{ad}(M[h])$ and $(^{ad}(M)) [h]$
are not isomorphic in $A \mbox{-{\rm gr}}$.
\end{rem}
\vskip 10pt

\section{Graded Cartan-Eilenberg Cohomology}
In this section we will extend the construction  in \cite{Ch} and
\cite{CE} to color Lie algebras.

\subsection{Color Kozsul Resolution}
Let $L$ be a $G$-graded $\varepsilon$-Lie algebra over $\K$, and let
$V$ be the underlying graded space of $L$. Set
\begin{align}
\wedge_{\varepsilon}V:=T(V)/\left\langle u\otimes
v+\varepsilon(u,v)v\otimes u\right\rangle
\end{align}
 where $u,v$ are homogeneous in $V$. Clearly $\wedge_\varepsilon V$
 is graded by the group $\mathbb{Z}\times G$, in other words,
 \begin{align}
 \wedge_\varepsilon V=\bigoplus_{n \geq 0} \wedge_\varepsilon^n V,
 \end{align}
 and each $\wedge_\varepsilon^n V$ is graded by $G$.\par \vskip 5pt

Define  $C_n:=U(L)\otimes_\K\wedge^n_{\varepsilon}V$, which  is
graded  by $G$ such that the degree of $u\otimes v$ is
 $|u|\cdot |v|$, for homogeneous  $u\in U(L)$ and $v\in\wedge^n_{\varepsilon}V$,
$n\in\N$. Endow $C_n$ with a left $U(L)$-module structure, which
is induced by the multiplication of $U(L)$. Obviously,  each $C_n$
is a graded $U(L)$-module (with respect to the group $G$) and it
is gr-free (again in the sense of \cite{NV}).\par \vskip 5pt

Denote by  $\left\langle x_1,...,x_n\right\rangle$  the
element $x_1\wedge\cdots\wedge x_n$ of $\wedge^n_{\varepsilon}V$.
Define, for every homogeneous $y\in L$, an $L$-module homomorphism
$\theta(y):C_n\rightarrow C_n$ by
\begin{align*}
\theta(y)(u\otimes\left\langle
x_1,...,x_n\right\rangle)&:=-\varepsilon(|y|,|u|)uy\otimes\left\langle
x_1,...,x_n\right\rangle\\
&+\sum_{i=1}^n\varepsilon(|y|,|u|\cdot |x_1|\cdots
|x_{i-1}|)u\otimes\left\langle
x_1,...,\left[y,x_i\right],...,x_n\right\rangle
\end{align*}
\par
\vskip 5pt
 We claim that
\begin{equation}\label{e1.1}
    \theta(x)\theta(y)-\varepsilon(|x|, |y|)\theta(y)\theta(x)=\theta([x,y])
\end{equation}
for all homogeneous $x,y,x_1,..,x_n\in L$ and $u\in U(L)$. We check
that $\theta$ verifies Eq.(\ref{e1.1}) and the fact that
$\theta=(\theta_1\otimes\theta_2)^{\chi}$ is the twisted tensor product of
$L$-module maps (see Section 3.4):
$$\theta_1(x):U(L)\rightarrow U(L), u\mapsto\theta_1(x)(u):=-\varepsilon(x,u)ux$$
and
$$\theta_2(x):\wedge^n L\rightarrow\wedge^n L, \left\langle x_1,...,x_n\right\rangle\mapsto\sum_i^n\varepsilon(x,x_i)\left\langle x_1,...,\left[x,x_i\right],...,x_n\right\rangle.$$

We define also, for every $y\in L$, a graded $L$-module homogeneous homomorphism of degree zero $\sigma(y):C_n\rightarrow C_{n+1}$ by
$$\sigma(y)(u\otimes\left\langle x_1,...,x_n\right\rangle):=\varepsilon(|y|,|u|)u\otimes\left\langle y,x_1,...,x_n\right\rangle$$
for all homogeneous elements $x_1,..,x_n\in L$ and $u\in U(L)$. It is easy to check that

\begin{equation}\label{e'1.2}
\sigma(\left[x,y\right])=\theta(x)\sigma(y)-\varepsilon(x,y)\sigma(y)\theta(x).
\end{equation}
Next we define by induction $L$-module homorphisms of degree zero $d_n:C_n\rightarrow C_{n-1}$ by
\begin{equation}\label{e1.3}
    \sigma(y)d_{n-1}+d_n\sigma(y)=-\theta(y)
\end{equation}
for all homogeneous elements $y\in L$ and $u\in U(L)$. We set $d_0:=0$. Since $u\otimes\left\langle x_1,...,x_n\right\rangle=\varepsilon(u,x_1)\sigma(x_1)(u\otimes\left\langle x_2,...,x_n\right\rangle)$, it follows from Eq. (\ref{e1.3}) that
\begin{align*}
d_n(u\otimes\left\langle x_1,...,x_n\right\rangle)&=\varepsilon(u,x_1)d_n\sigma(x_1)(u\otimes\left\langle x_2,...,x_n\right\rangle)\\
&=\varepsilon(u,x_1)(-\theta(x_1)-\sigma(x_1)d_{n-1})(u\otimes\left\langle x_2,...,x_n\right\rangle).\\
&
\end{align*}
We deduce that the operator $d_n$ is explicitly given  by
\begin{align*}
&d_n(u\otimes\left\langle x_1, \cdots, x_{n}\right\rangle)\\
&=\sum_{i=1}^{n} (-1)^{i+1} \varepsilon_i\;ux_i\otimes\left\langle x_1, \cdots, \hat{x_i}, \cdots, x_{n}\right\rangle\\
&\quad +\sum_{1\leq i< j \leq n} (-1)^{i+j} \varepsilon_i \varepsilon_j\varepsilon(|x_j|,|x_i|)\;u\otimes\left\langle [x_i, x_j], x_1, \cdots, \hat{x_i}, \cdots, \hat{x_j}, \cdots, x_{n}\right\rangle,
\end{align*}
for all homogeneous elements $u \in U(L)$ and $x_i \in L$, with $\varepsilon_i=\prod_{h=1}^{i-1}\varepsilon(|x_h|, |x_i|)$ $i\geq 2$, $\varepsilon_1=1$ and the sign\quad$\widehat{}$\quad indicates that the element below it must be omitted.
We will show preceding by induction on $n\in\N$ that
\begin{equation}\label{e1.5}
    \theta(y)d_n=d_n\theta(y)
\end{equation}
It is obvious if $n=0$. For $n\geq 1$, we have
\begin{align*}
&\theta(y)d_n-d_n\theta(y)(u\otimes\left\langle x_1,...,x_n\right\rangle)\\
&=\epsilon(u,x_1)\left(\theta(y)d_{n}\sigma(x_1)-d_n\theta(y)\sigma(x_1)\right)(u\otimes\left\langle x_2,...,x_n\right\rangle).
\end{align*}
Since $\varepsilon(u,x_1)\neq 0$ for all homogeneous elements $x_1,u$, it sufficient to show that
$$\theta(y)\delta_{n}\sigma(x)-d_n\theta(y)\sigma(x)=0.$$
On the other hand\\
$\theta(y)d_{n}\sigma(x)-d_n\theta(y)\sigma(x)$\\
$=-\theta(y)\theta(x)-\theta(y)\sigma(x)d_{n-1}-\varepsilon(y,x)d_n\sigma(x)\theta(y)-d_n\sigma(\left[y,x\right])$, (by Eq. (\ref{e1.3}) and Eq.(\ref{e'1.2}))\\
$=-\theta(y)\theta(x)-\theta(y)\sigma(x)d_{n-1}+\varepsilon(y,x)(\theta(x)\theta(y)+\sigma(x)d_{n-1}\theta(y))+\theta\left[y,x\right]$\\
$+\sigma\left[y,x\right]d_{n-1})$, (by (\ref{e1.3}))\\
$=\left\{-\theta(y)\theta(x)+\varepsilon(y,x)\theta(x)\theta(y)+\theta\left[y,x\right]\right\}$\\
+$\left\{-\theta(y)\sigma(x)+\varepsilon(y,x)\sigma(x)\theta(y)+\sigma\left[y,x\right]\right\}d_{n-1}$, (by the induction hypothesis)\\
$=0$, (by Eq. (\ref{e1.1}), and Eq. (\ref{e'1.2})).
\par \vskip 5pt

Finally, we show that
\begin{equation}
    d_{n-1}d_{n}=0.
\end{equation}
It is obvious that $d_0d_1=0$. We reason by induction. We have, for $n\geq 2:$
$$d_{n-1}d_{n}(u\otimes\left\langle x_1...x_n\right\rangle)=\varepsilon(u,x_1)d_{n-1}d_{n}\sigma(x_1)(u\otimes\left\langle x_2...x_n\right\rangle),$$
from Eq. (\ref{e1.3}) we obtain:
\begin{align*}\varepsilon(u,x_1)d_{n-1}d_{n}\sigma(x_1)(u\otimes\left\langle x_2...x_n\right\rangle)&=-\varepsilon(u,x_1)d_{n-1}(\theta(x_1)+\sigma(x_1)d_{n-1})\\
&=\varepsilon(u,x_1)d_{n-1}d_{n}\sigma(x_1)(u\otimes\left\langle x_2...x_n\right\rangle)\\
&=-\varepsilon(u,x_1)d_{n-1}(\theta(x_1)+\sigma(x_1)d_{n-1})\\
&=-\varepsilon(u,x_1)(d_{n-1}\theta(x_1)+d_{n-1}\theta(x_1)\\
&+\sigma(x_1)d_{n-2}d_{n-1})\\
&=0
\end{align*}
from Eq. (\ref{e1.5}) and the induction hypothesis.\par \vskip5pt

Let $\left\{x_i\right\}_I$ be a homogeneous basis of $L$, where
$I$ is a well-ordered set. By the generalized PBW theorem the
elements
\begin{equation}\label{e113}
    e_{k_1}\cdots e_{k_m}\otimes\left\langle e_{l_1}\cdots e_{l_n}\right\rangle
\end{equation}
with
\begin{equation}
    k_1\leq\cdots\leq k_m\quad\mathrm{and}\quad k_i<k_{i+1}\quad\mathrm{if}\quad\varepsilon(|e_{k_i}|,|e_{k_i}|)=-1
\end{equation}
and
\begin{equation}
    l_1\leq\cdots\leq l_n\quad\mathrm{and}\quad l_i<l_{i+1}\quad\mathrm{if}\quad\varepsilon(|e_{l_i}|,|e_{l_i}|)=1
\end{equation}
form a homogeneous basis of $C_n$.\par \vskip 5pt

We define a family of $G$-graded subspace $F_pC$ of $C$ with $p\in\Z$, as follows: $(F_pC)_{-1}:=\K$ and $(F_pC)_n$, $n\geq 0$, is the subspace of $C_n$ generated by the homogeneous basis (\ref{e113}) with $m+n\leq p$.
 We see that for all $n\geq 0$ the differential $d_n$ maps $(F_pC)_n$ into $(F_pC)_{n-1}$,
 then $F_pC$ is a $G$-graded subcomplex of $C$.
 For every $p\geq 1$ we define a $G$-graded complex $W^p$ by $(F_pC)_n/(F_pC)_{n-1}$ for $n\geq 0$
 and $W^p_{-1}:=\K$. It is now clear that the differential $d^p_n:W^p_n\rightarrow W^{p}_{n-1}$ is $G$-graded
  and given by\vskip 5pt
\noindent $d^p_n(e_{k_1}\cdots e_{k_m}\otimes\left\langle
e_{l_1}\cdots e_{l_n}\right\rangle)\equiv$
$$ \sum_{i=1}^{n} (-1)^{i+1}\prod_{h=1}^{i-1}\varepsilon(|e_{l_h}|,|e_{l_i}|)\;e_{k_1}\
\cdots e_{k_m}e_{l_i}\otimes\left\langle e_{l_1}, \cdots, \hat{e_{l_i}}, \cdots, e_{l_n}
\right\rangle\mathrm {mod}(F_{p-1}C)_{n-1}.$$
Note that the summands on the right hand side are not necessarily
of the form (\ref{e113}), since we cannot guarantee $k_m\leq l_i$.
\par
\vskip 5pt

\begin{lem}\label{n11}
 We have that the homology groups $\mathrm{H}_n(W^p)=0$ for all $p\geq 1$ and all $n$.
\end{lem}
\begin{proof} We define a $G$-graded homomorphisms $t^p_n$ as follows: $t^p_{-1}:\K\rightarrow W^p_0$ is given by $t^p_{-1}(1_{\K}):=1\otimes\left\langle \right\rangle$ and, for $n\geq 0$, we define $t^p_n:W^p_n\rightarrow W^{p}_{n+1}$ by
$$t^p_n(e_{k_1}\cdots e_{k_m}\otimes\left\langle e_{l_1}\cdots e_{l_n}\right\rangle)$$
$$\equiv\sum_{i=1}^{m} \prod_{h=i+1}^{m}\varepsilon|e_{k_i}|,|e_{k_h}|)\;e_{k_1}\cdots\hat{e_{k_i}}\cdots e_{k_m}\otimes\left\langle e_{k_i},e_{l_1},\cdots, e_{l_n}\right\rangle\mathrm {mod}(F_{p-1}C)_{n}.$$
We will show that
$$d^{p}_{n+1}t^p_n+t^{p}_{n-1}d_n^p=p\;\mathrm{id}.$$

\begin{align*}
&d^{p}_{n+1}t^p_n(e_{k_1}\cdots e_{k_m}\otimes\left\langle e_{l_1}\cdots e_{l_n}\right\rangle)\\
&=\sum_{i,j=1}^{m,n} \; (-1)^{j}\prod_{h=i+1}^{m}\varepsilon(|e_{k_i}|,|e_{k_h}|)\; \varepsilon(|e_{k_i}|,|e_{l_j}|)\; \prod_{h=1}^{j-1}\varepsilon(|e_{l_h}|,|e_{l_j}|)\\
&\qquad \qquad  e_{k_1}\cdots\hat{e_{k_i}}\cdots e_{k_m}e_{l_j}\otimes\left\langle e_{k_i},e_{l_1},\cdots,\hat{e_{l_j}}\cdots, e_{l_n}\right\rangle\\
&\quad
 +\sum_{i=1}^m(-1)^{1+1}\prod_{h=i+1}^{m}\varepsilon(|e_{k_i}|,|e_{k_h}|)\; e_{k_1}\cdots\hat{e_{k_i}}\cdots
e_{k_m}e_{k_i}\otimes\left\langle e_{l_1}\cdots
e_{l_n}\right\rangle\\
&=\sum_{i,j=1}^{m,n} \; (-1)^{j}\prod_{h=i+1}^{m}\varepsilon(|e_{k_i}|,|e_{k_h}|)\; \varepsilon(|e_{k_i}|,|e_{l_j}|)\; \prod_{h=1}^{j-1}\varepsilon(|e_{l_h}|,|e_{l_j}|)\\
&\qquad \qquad  e_{k_1}\cdots\hat{e_{k_i}}\cdots e_{k_m}e_{l_j}\otimes\left\langle e_{k_i},e_{l_1},\cdots,\hat{e_{l_j}}\cdots, e_{l_n}\right\rangle\\
&\quad +\sum_{i=1}^m
\prod_{h=i+1}^{m}\varepsilon(|e_{k_i}|,|e_{k_h}|)\;
\prod_{h=i+1}^{m}\varepsilon(|e_{k_h}|,|e_{k_i}|)\; e_{k_1}\cdots
e_{k_i}\cdots e_{k_m}\otimes\left\langle e_{l_1}\cdots
e_{l_n}\right\rangle\\
&=\sum_{i,j=1}^{m,n} \; (-1)^{j}\prod_{h=i+1}^{m}\varepsilon(|e_{k_i}|,|e_{k_h}|)\; \varepsilon(|e_{k_i}|,|e_{l_j}|)\; \prod_{h=1}^{j-1}\varepsilon(|e_{l_h}|,|e_{l_j}|)\\
&\qquad \qquad  e_{k_1}\cdots\hat{e_{k_i}}\cdots e_{k_m}e_{l_j}\otimes\left\langle e_{k_i},e_{l_1},\cdots,\hat{e_{l_j}}\cdots, e_{l_n}\right\rangle\\
& \quad +m\; e_{k_1}\cdots e_{k_i}\cdots e_{k_m}\otimes\left\langle
e_{l_1}\cdots e_{l_n}\right\rangle.
\end{align*}
(The last equality uses the fact that
$\prod_{h=i+1}^{m}\varepsilon(|e_{k_i}|,|e_{k_h}|)\;
\prod_{h=i+1}^{m}\varepsilon(|e_{k_h}|,|e_{k_i}|)=1$.)\par \vskip
5pt

And we have\\
\begin{align*}
&t^p_{n-1}d^{p}_{n}(e_{k_1}\cdots e_{k_m}\otimes\left\langle e_{l_1}\cdots e_{l_n}\right\rangle))\\
&=\sum_{i,j=1}^{m,n} (-1)^{j+1}\prod_{h=1}^{j-1}\; \varepsilon(|e_{l_h}|,|e_{l_j}|)\; \prod_{h=i+1}^{m-1}\varepsilon(|e_{k_i}|,|e_{k_h}|)\;\varepsilon(|e_{k_i}|,|e_{l_j}|)\\
&\qquad \qquad e_{k_1}\cdots\hat{e_{k_i}}\cdots
e_{k_m}e_{l_j}\otimes\left\langle
e_{k_i},e_{l_1},\cdots,\hat{e_{l_j}}\cdots, e_{l_n}\right\rangle\\
&\quad
+\sum_{j=1}^n(-1)^{j+1}\prod_{h=1}^{j-1}\varepsilon(|e_{l_h}|,|e_{l_j}|)\;
e_{k_1}\cdots e_{k_m}e_{l_j}\otimes\left\langle
e_{l_1},\cdots,\hat{e_{l_j}}\cdots,
e_{l_n}\right\rangle\\
&=-\sum_{i,j=1}^{m,n} (-1)^{j}\prod_{h=1}^{j-1}\varepsilon(|e_{l_h}|,|e_{l_j}|)\; \prod_{h=i+1}^{m-1}\varepsilon(|e_{k_i}|,|e_{k_h}|)\; \varepsilon(|e_{k_i}|,|e_{l_j}|)\\
&\qquad \qquad e_{k_1}\cdots\hat{e_{k_i}}\cdots
e_{k_m}e_{l_j}\otimes\left\langle
e_{k_i},e_{l_1},\cdots,\hat{e_{l_j}}\cdots, e_{l_n}\right\rangle\\
&\quad + \sum_{j=1}^n(-1)^{j+1}(-1)^{j+1}\; \prod_{h=1}^{j-1}\
\varepsilon(|e_{l_h}|,|e_{l_j}|)\;
\prod_{h=1}^{j-1}\varepsilon(|e_{l_j}|,|e_{l_h}|)\; e_{k_1}\cdots
e_{k_m}\otimes\left\langle e_{l_1},\cdots ,e_{l_j}\cdots,
e_{l_n}\right\rangle\\
&=-\sum_{i,j=1}^{m,n} (-1)^{j}\prod_{h=1}^{j-1}\varepsilon(|e_{l_h}|,|e_{l_j}|)\; \prod_{h=i+1}^{m-1}\varepsilon(|e_{k_i}|,|e_{k_h}|)\; \varepsilon(|e_{k_i}|,|e_{l_j}|)\\
&\qquad \qquad e_{k_1}\cdots\hat{e_{k_i}}\cdots
e_{k_m}e_{l_j}\otimes\left\langle
e_{k_i},e_{l_1},\cdots,\hat{e_{l_j}}\cdots,
e_{l_n}\right\rangle\\
&\quad + n\; e_{k_1}\cdots e_{k_m}\otimes\left\langle
e_{l_1},\cdots, e_{l_j}\cdots, e_{l_n}\right\rangle.
\end{align*}
(Here again we use
$\prod_{h=1}^{j-1}\varepsilon(|e_{l_h}|,|e_{l_j}|)\;
\prod_{h=1}^{j-1}\varepsilon(|e_{l_j}|,|e_{l_h}|)=1$.) \vskip 3pt
From that we deduce that
\begin{equation}
    (d^{p}_{n+1}t^p_n+t^{p}_{n-1}d_n^p)\; (e_{k_1}\cdots e_{k_m}\otimes\left\langle e_{l_1}\cdots e_{l_n}\right\rangle)=(m+n)(e_{k_1}\cdots e_{k_m}\otimes\left\langle e_{l_1}\cdots
    e_{l_n}\right\rangle),
\end{equation}
We set
\begin{equation}
    t^p=\bigoplus_{n,m\in\N,n+m=p\neq 0}\frac{1}{p}t^{p}_n,
\end{equation}
and thus, we deduce that
\begin{equation}
    d^pt^p+t^pd^p=\mathrm{Id}.
\end{equation}
 Hence $\mathrm{H}_n(W^p)=0$ for all $p\geq 1$ and all $n$.
\end{proof}

\par \vskip 10pt

The following theorem gives the color Koszul resolution.
\begin{thm}\label{t3}
Then the sequence
\begin{equation}
    C:\quad \cdots \rightarrow
    C_n\stackrel{d_n}{\rightarrow}C_{n-1}\rightarrow\cdots \rightarrow C_1\stackrel{\epsilon}{\rightarrow}C_0
    \rightarrow 0
\end{equation}
is a $G$-graded $U(L)$-free resolution of the $G$-graded trivial
module $\K$ via $\epsilon$.
\end{thm}
\vskip 2pt

\begin{proof} We consider the ($G$-graded) exact sequence of complexes
\begin{equation}
    0\rightarrow F_{p-1}\rightarrow F_pC\rightarrow W^p.
\end{equation}
For the associated long exact homology sequence it follows from Lemma \ref{n11} that

\begin{equation}
    \mathrm{H}_n(F_{p-1}C)\simeq \mathrm{H}_n(F_pC)
\end{equation}
for all $n$, and all $p\geq 1$. Since $F_0C$ is the graded complex
$0\rightarrow\K\rightarrow\K\rightarrow 0$, we then obtain
$\mathrm{H}_n(F_0C)=0$, for all $n$. Hence, by induction,
$\mathrm{H}_n(F_pC)=0$ for all $n$ and all $p\geq 0$. Since
$C=\cup_{p\geq 0}F_pC$ then the result follows that
$\mathrm{H}_n(F_pC)=0$.
\end{proof}

\subsection{Cohomology of color Lie Algebras}
Let $M$ be a left $L$-module, we define the $n$-th graded
cohomology group of $L$ with value in $M$ as
\begin{equation}
    \mathrm{H}^n(L,M)_h:=\mathrm{EXT}_{U(L)}^n(\K,M)_h=\mathrm{Ext}_{U(L)\mbox{-gr}}^n(\K,M[h]).
\end{equation}
for all $h\in G$, where $\K$ is the trivial graded $L$-module, or
equivalently, $U(L)$-module.\par

We also define
\begin{equation}
\mathrm{H}^n_{\rm gr}(L,M)=\mathrm{H}^n(L,M)_e.
\end{equation}
Thus $\mathrm{H}^n(L,M)_h= \mathrm{H}^n_{\rm gr}(L,M[h]) $.\par
\vskip 5pt
Set
$$\mathrm{H}(L,M)=\oplus_{h\in G}\mathrm{H}^n(L,M)_h.$$


To compute ${\mathrm H}^n_{\rm
gr}(L, M)$, we may be used the gr-free resolution of the
trivial module $\K$ in Theorem \ref{t3}. Let $M$ be a graded $L$-module, the
cohomology groups ${\mathrm H}^n_{\rm
gr}(L, M)$ are the homology groups of the complex 
\begin{align*}
\Hom_{U(L)\mbox{-gr}}(C_n,M)&=\Hom_{U(L)\mbox{-gr}}(U(L)\otimes\wedge^n_{\varepsilon}L,M)\\
                            &\simeq \Hom_{\rm gr} (\wedge^n_{\varepsilon}L,M),
\end{align*}
where $C$ is the complex in Theorem \ref{t3}.
Under the above isomorphisms, the corresponding differential
operator is given by
\begin{align}
&\delta^n(f)\left(x_1, \cdots, x_{n+1}\right)\\
&=\sum_{i=1}^{n+1} (-1)^{i+1}\varepsilon_i\;x_i\cdot f\left( x_1, \cdots, \hat{x_i}, \cdots, x_{n+1}\right)\\
&\quad +\sum_{1\leq i< j \leq n+1} (-1)^{i+j}\varepsilon_i
\varepsilon_j\varepsilon(|x_j|,|x_i|)\;f\left( [x_i, x_j], x_1,
\cdots, \hat{x_i}, \cdots, \hat{x_j}, \cdots, x_{n+1}\right),
\end{align}
for all $f \in  \Hom_{\rm gr} (\wedge^n_{\varepsilon}L,M)$, where
the $\varepsilon_i$'s are given in 4.1.  This description of the
graded cohomology groups $\mathrm{H}^n_{\rm gr}(L,B)$ shows that
these coincide,  in the case of degree $e$, with the graded
Cartan-Eilenberg cohomology of $L$ introduced by Scheunert and
Zhang in(\cite{S2},\cite{SZ}).\par \vskip 5pt

Now we can apply Corollary \ref{c1} to the universal enveloping
algebra $U(L)$ of a $G$-graded $\varepsilon$-Lie algebra $L$: by
the Example in Section 3, we see $U(L)$ is a color Hopf algebra.
Note that if $M$ is a graded  $U(L)$-bimodule, the corresponding
adjoint $L$-module $^{ad}(M)$ is given by (compare (\ref{e2.5}))
\begin{align}\label{e111}
xm= x.m-\varepsilon(|x|, |m|)m.x
\end{align}
for homogeneous $x\in L$ and $m \in M$. \par \vskip 10pt
In summary, we get
\begin{thm}\label{t5}
 Let $L$ be a $G$-graded $\varepsilon$-Lie algebra, and
 let $U(L)$ be its universal enveloping algebra.
 Let $M$  be a  graded $U(L)$-bimodule. Then there exists an isomorphism of graded spaces
 \begin{equation}
 \mathrm{HH}^n(U(L),M)_h\simeq\mathrm{H}^n(L,^{ad}(M))_h=\mathrm{H}_{\rm gr}^n(L,^{ad}(M[h])). \quad
 n\geq 0.
 \end{equation}
 In particular we obtain
 \begin{equation}
 \mathrm{HH}^n_{\rm gr}(U(L),M)\simeq\mathrm{H}_{\rm gr}^n(L,^{ad}(M)) , \quad
 n\geq 0.
 \end{equation}
 \end{thm}

\end{document}